\documentclass[a4,11pt]{cip-v3}
\usepackage{mathptmx,siunitx}
\usepackage{mathptmx}
\usepackage[T1]{fontenc}
\DeclareSymbolFont{epsilon}{OML}{cmm}{m}{it}
\DeclareMathSymbol{\epsilon}{\mathord}{epsilon}{"0F}
\usepackage[colorlinks]{hyperref}
\hypersetup{
	colorlinks = true,
	linkcolor = blue,
	anchorcolor = blue,
	citecolor = blue,
	filecolor = blue,
	urlcolor = blue
}
\usepackage{graphics,graphicx,subfigure,wrapfig,epstopdf}
\usepackage{longtable}

\usepackage{wasysym,amsmath,amsxtra,amssymb,latexsym,amscd,color,cite,bm,array,multirow}
\usepackage[mathscr]{eucal}
\usepackage{bbold}
\def\authors#1{\author{\begin{flushleft}{#1}\end{flushleft}}}
\def\authord#1#2{\textbf{\indent{#1}$^{#2}$}}
\def\addressed#1#2{\\[1mm]\textit{$\!\!\!^{#1}$\indent#2}}
\def\CorrEmail#1{\\[4mm]
	\textit{E-mail:}~$^\dag${#1}}

\def\PublicationInformation#1#2#3{\\[3mm]\textit{\indent Received~#1}\\[1mm]
	\textit{Accepted for publication~#2}\\[1mm]
	\textit{Published~#3}}

\def\Keywords#1{$\qquad$\\[-.35cm] \textnormal{Keywords:~{#1}}.} 

\def\and{\textbf{and} }
\def\Classification#1{$\quad$\\[-.35cm] \textnormal{Classification numbers:~{#1}.}}

\usepackage{physics}
\usepackage{booktabs}
\usepackage{physics}
\usepackage{mhchem}
\usepackage{cases}
\begin{document}
\Year{2026}
	\title{A Direct Algebraic Approach to Normal Ordering of Exponential Bosonic Operators with Applications to  Two-Dimensional Excitonic Form Factors}
	\authors{
	\authord{Duy-Anh P. Nguyen}{1}, \authord{Ngoc-Tram D. Hoang}{2}, \authord{Dang-Khoa D. Le}{2}, \authord{Van-Hoang Le}{2,\dagger}
	\newline
	\newline
	\addressed{1}{Thu Dau Mot University, Phu Loi, Ho Chi Minh City 75110, Vietnam}
	\addressed{2}{Computational Physics Key Laboratory K002, Department of Physics, Ho Chi Minh City University of Education, Ho Chi Minh City 72759, Vietnam}
\CorrEmail{hoanglv@hcmue.edu.vn}
\PublicationInformation{\today}{(Editor will insert the date when the article is accepted)}{(Editor will insert the date when the article is published)}
}
\maketitle
	\markboth{Title of the article $\ldots$}{First Author \textit{et al.}}

\begin{abstract}
We develop a systematic algebraic approach, based on the Wei--Norman factorization method, to the normal ordering of exponential bosonic operators and apply it to derive analytical excitonic form factors in two-dimensional semiconducting materials. By introducing an auxiliary parameter, the normal-ordering problem is reduced to a system of ordinary differential equations determined by the commutation relations of the underlying closed Lie algebra. The approach is first illustrated for exponential operators associated with the Heisenberg--Weyl and $su(1,1)$ algebras, and is then extended to two-mode bosonic operators involving $su(2)$ and a six-generator closed algebra that contains two coupled $su(1,1)$ subalgebras. For the excitonic application, the Levi--Civita transformation maps the two-dimensional exciton problem onto an oscillator representation, providing a natural formulation in terms of bosonic creation and annihilation operators. Combined with the Laplace and Fourier representations of the Rytova--Keldysh potential, this formulation reduces the interaction matrix elements to the evaluation of exponential bosonic form factors. The isotropic problem is governed by a three-generator $su(1,1)$ algebra, whereas the anisotropic case requires the full six-generator algebra together with an additional $su(2)$ factorization. Explicit analytical expressions for both form factors, $\langle e^{-rt}\rangle$ and $\langle e^{i\mathbf q\cdot\mathbf r}\rangle$, are obtained, providing useful building blocks for matrix-element calculations in two-dimensional excitonic systems and potentially in other quantum problems involving exponential bosonic operators.

\end{abstract}


\Classification{02.20.Uw, 03.65.Fd, 42.50.-p, 71.35.-y, 03.65.Ca}

\section{Introduction}

The normal ordering of bosonic operators is a central technique in quantum mechanics and quantum field theory. By rearranging creation and annihilation operators in prescribed order, complicated operator expressions can be transformed into forms suitable for analytical evaluation of matrix elements, expectation values, and correlation functions. Consequently, normal-ordering techniques have found widespread applications in quantum optics, many-body theory, quantum statistical mechanics, and condensed matter physics~\cite{glauber1963, cahill1969, louisell1973, mandel1995}. While finite polynomials of bosonic operators can usually be treated by repeated application of the canonical commutation relations, exponential bosonic operators present a considerably greater challenge because their normal ordering generally requires the factorization of exponential operators involving noncommuting generators, a process traditionally referred to as operator disentangling. Therefore, the normal ordering of exponential bosonic operators has remained a longstanding and important problem in mathematical physics.

A powerful approach to the factorization of exponential operators originates from the classical works of Campbell~\cite{campbell1896}, Baker~\cite{baker1905}, and Hausdorff~\cite{hausdorff1906}, which established the Baker--Campbell--Hausdorff (BCH) formula relating products of exponential operators to a single exponential. Although mathematically exact, the BCH expansion rapidly becomes cumbersome due to the infinite hierarchy of higher-order commutators. An important advance was made by Wilcox~\cite{wilcox1967}, who developed a differential formulation for parameter-dependent exponential operators. Building upon these developments, Wei and Norman~\cite{wei1963, wei1964} proposed a general factorization procedure for exponential operators associated with finite-dimensional Lie algebras, which reduces the determination of a factorized representation to the solution of a system of ordinary differential equations.  Because of its generality and its reliance solely on the commutation relations of the underlying Lie algebra, the Wei--Norman formalism has become a standard analytical tool with applications in quantum optics, coherent and squeezed states, time-dependent quantum systems, quantum control, and Lie-group integration \cite{truax1985, ban1993, rau2000, gerry2000, altafini2002, charzynski2013}.

Although the Wei--Norman formalism provides a general framework for the factorization of exponential operators associated with Lie algebras, its potential for constructing normal-ordered representations of exponential bosonic operators has not been formulated within a unified algebraic framework. The key observation is that, by selecting an appropriate ordering of the Lie-algebra generators, the factorized expression obtained from the Wei--Norman procedure can be brought directly into the bosonic normal ordering, with creation operators appearing to the left of annihilation operators. This establishes a natural connection between Lie-algebraic factorization and bosonic normal ordering. Building on this idea, the present paper develops a systematic algebraic framework for constructing normal-ordered representations of exponential bosonic operators that arise in the analytical evaluation of excitonic form factors in two-dimensional materials.

The application of bosonic operator techniques to two-dimensional excitonic systems traces its origin to the Levi--Civita transformation, which maps the two-dimensional hydrogenic problem onto a two-dimensional harmonic oscillator. In our earlier work~\cite{nguyen1993}, this transformation established a natural bosonic representation of the excitonic Hamiltonian, enabling the Feranchuk--Komarov (FK) operator method~\cite{feranchuk1982, feranchuk2015} to be applied to both analytical and numerical studies of two-dimensional excitons \cite{tram2013, nam2017, nguyen2019, ly2023}. Within this framework, excitonic wave functions are represented in the bosonic Fock space, while physical observables are expressed as matrix elements of bosonic operators. Consequently, the evaluation of excitonic form factors becomes one of the central problems of the operator formalism. Using the Fourier and Laplace representations of the Rytova--Keldysh potential~\cite{rytova1967, keldysh1979}, the evaluation of these form factors is naturally reduced to the calculation of matrix elements of exponential bosonic operators, thereby providing a direct connection between excitonic calculations and the normal-ordering problem addressed in the present work.

Motivated by the analytical evaluation of excitonic form factors, we have progressively developed an analytical method for constructing normal-ordered representations of exponential bosonic operators associated with different representations of the Rytova--Keldysh potential. Our initial study of isotropic excitons employed the Fourier representation of the interaction, leading to exponential operators governed by a six-generator bosonic Lie algebra~\cite{nguyen2019}. Owing to the rotational symmetry of the isotropic system, the excitonic Schr\"odinger equation is separable in polar coordinates, with the angular dependence described by the functions $e^{-im\theta}$. Consequently, the angular part is eliminated analytically in the evaluation of the form factors, reducing the matrix-element calculation effectively to the $m=0$ case. Subsequently, we demonstrated that, for isotropic systems, the Laplace representation of the Rytova--Keldysh potential provides a more direct analytical formulation, reducing the problem to a three-generator Lie algebra and yielding compact analytical expressions for the excitonic form factors~\cite{ly2023}. 

More recently, our studies have been extended to anisotropic two-dimensional materials~\cite{le2026cip, le2026prb}, where the Fourier representation again becomes necessary. In this case, the rotational symmetry is broken, the angular and radial motions are no longer separable, and the simplifications associated with the isotropic case disappear. Although the underlying six-generator Lie algebra remains unchanged, the coupled angular and radial motions make the construction of normal-ordered representations and the evaluation of the corresponding form factors substantially more involved, thereby necessitating a systematic algebraic treatment. The present work establishes such a unified framework, encompassing both the three-generator and six-generator Lie algebras that arise in the analytical evaluation of excitonic form factors based on the Laplace and Fourier representations of the Rytova--Keldysh interaction.

The rest of this paper is organized as follows. Section~\ref{sec:2} develops the algebraic approach to normal ordering for polynomial and
exponential bosonic operators associated with the Heisenberg--Weyl and $su(1,1)$ algebras. Section~\ref{sec:3} extends the approach to
two-mode bosonic operators and applies it to the analytical evaluation of excitonic form factors for isotropic and anisotropic two-dimensional
systems. Finally, Sec.~\ref{sec:4} summarizes the main results and conclusions.

\section{Algebraic Approach to Normal Ordering of Bosonic Operators}
\label{sec:2}

The normal ordering of bosonic creation and annihilation operators is a standard topic in quantum mechanics and quantum optics and is discussed
in many textbooks~\cite{gerry2005, walls2008, feranchuk2015}. In this section, we first briefly review the normal ordering of polynomial
bosonic operators to establish the notation and basic algebraic procedure. We then extend this procedure to exponential operators, where the normal-ordering problem is formulated in terms of the Wei--Norman factorization method for the corresponding closed Lie algebras.

\subsection{Normal Ordering of Polynomial Bosonic Operators}
\label{subsec:2.1}

Let $\hat a$ and $\hat a^\dagger$ be bosonic annihilation and creation operators satisfying the canonical commutation relation
\begin{equation}\label{eq1}
[\hat a,\hat a^\dagger]=1.
\end{equation}
The problem of normal ordering consists in rewriting operator products such that all creation operators $\hat a^{\dagger}$ appear to the left of annihilation operators $\hat a$. From Eq.~\eqref{eq1}, one obtains the basic identity
\begin{equation}\label{eq2}
\hat a \hat a^{\dagger} = \hat a^{\dagger} \hat a + 1,
\end{equation}
which allows one to systematically interchange the order of operators.

We first consider products of the form $\hat a(\hat a^{\dagger})^n$, for which Eq.~\eqref{eq2} gives the simplest case, $n=1$. Applying Eq.~\eqref{eq2} repeatedly, one obtains for $n=2$:
\begin{eqnarray}
\hat a (\hat a^{\dagger})^2&=& (\hat a{\hat a^{\dagger}}){\hat a^{\dagger}}= ({\hat a^{\dagger}}\hat a+1){\hat a^{\dagger}}=
{\hat a^{\dagger}}( {\hat a^{\dagger}}{\hat a}+1)+{\hat a^{\dagger}}
=(\hat a^{\dagger})^2 \hat a + 2 \hat a^{\dagger}.\nonumber
\end{eqnarray}
This result suggests the general normal-ordering formula
\begin{equation}\label{eq3}
\hat a (\hat a^{\dagger})^n=
(\hat a^{\dagger})^n \hat a + n(\hat a^{\dagger})^{n-1},
\end{equation}
which follows straightforwardly by induction.

We now turn to the general product $\hat a^m (\hat a^\dagger)^n$. Equation \eqref{eq3} gives the normal-ordered form for the case of $m=1$.  We first compute the cases of $m=2$ and $3$ by repeatedly applying Eq.~\eqref{eq3}, which allows one to move each annihilation operator through the string of creation operators. The results
\begin{eqnarray}
\hat a^2 (\hat a^\dagger)^n
&=&(\hat a^\dagger)^n \hat a^2+2n (\hat a^\dagger)^{n-1} {\hat a} +n(n-1)(\hat a^\dagger)^{n-2},\nonumber\\
\hat a^3 (\hat a^\dagger)^n
&=& (\hat a^\dagger)^n \hat a^3+3n(\hat a^\dagger)^{n-1} \hat a^2
+3n(n-1)(\hat a^\dagger)^{n-2} \hat a+n(n-1)(n-2)(\hat a^\dagger)^{n -3},\nonumber
\end{eqnarray}
lead naturally to the general normal-ordering formula
\begin{equation}\label{eq4}
\hat a^m(\hat a^\dagger)^n=\sum_{k=0}^{\min(m,n)}
\binom{m}{k}\binom{n}{k}k! (\hat a^\dagger)^{n-k}\hat a^{m-k},
\end{equation}
where $\binom{n}{k}= n!/(n-k)!k!$ denotes the binomial coefficient. Equation~\eqref{eq4} can be established rigorously by induction on $m$.

The structure of Eq.~\eqref{eq4} shows that normal ordering is governed by a simple combinatorial mechanism: each commutation produces either
a reordered term or a scalar contribution. For exponential operators, however, repeated commutation becomes increasingly cumbersome, motivating
a different algebraic treatment based on the closed Lie algebra generated by the operators in the exponent.

\subsection{The Heisenberg--Weyl Algebra and Normal Ordering of Exponentials Linear in Bosonic Operators }
\label{subsec:2.2}

As a first illustration of the method, we consider the exponential operator $e^{\hat a^\dagger+\hat a}$, whose exponent is linear in the
bosonic creation and annihilation operators. Its normal-ordered form is known from the Baker--Campbell--Hausdorff (BCH) formula
\cite{baker1905, truax1985} as
\begin{equation}\label{eq5}
\hat A=e^{\hat a^\dagger + \hat a}=e^{1/2} e^{\hat a^\dagger} e^{\hat a}.
\end{equation}
Although this result is standard, we rederive it here as a simple illustration of the general procedure developed in this work. In particular, this example demonstrates how the normal-ordering problem for exponential operators can be reduced to a system of ordinary differential equations.

To this end, we introduce an auxiliary parameter $t$ and define
\begin{equation}\label{eq6}
\hat A(t) = e^{t(\hat a^\dagger + \hat a)},
\end{equation}
from which it follows that $\hat A = \hat A(1)$.

Since the operators $(\hat a^\dagger, \hat a, 1)$ form the Heisenberg--Weyl algebra, we assume that $\hat A(t)$ can be written in the factorized form
\begin{equation}\label{eq7}
\hat A(t) = e^{f(t)} e^{h(t)\hat a^\dagger} e^{g(t)\hat a},
\end{equation}
where $f(t)$, $h(t)$, and $g(t)$ are scalar functions satisfying the initial conditions
\begin{equation}\label{eq8}
f(0)=h(0)=g(0)=0.
\end{equation}

Differentiating both sides of Eq.~\eqref{eq7}, and then multiplying the resulting equation from the right by $\hat A^{-1}(t)$ yields the operator equation
\begin{equation}\label{eq9}
\hat a^\dagger + \hat a = f'(t)+h'(t)\hat a^\dagger
+g'(t)\,e^{h(t)\hat a^\dagger}\hat a e^{-h(t)\hat a^\dagger}.
\end{equation}
Furthermore, using the BCH formula, we obtain
\begin{equation}
e^{h\hat a^\dagger}\hat a e^{-h\hat a^\dagger}
=\hat a - h\,, \nonumber
\end{equation}
which simplifies Eq.~\eqref{eq9} and yields the system of differential equations
\begin{equation}\label{eq10}
h'(t)=1, \qquad g'(t)=1,\qquad f'(t)=h(t).
\end{equation}
With the initial conditions \eqref{eq8}, the solutions are
\begin{equation}\label{eq11}
h(t)=t,\qquad g(t)=t,\qquad f(t)=\frac{t^2}{2}\,.
\end{equation}
Substituting these solutions into Eq.~\eqref{eq7} and setting $t=1$, we obtain the normal-ordered form
\begin{equation}
e^{\hat a^\dagger + \hat a}=e^{1/2}e^{\hat a^\dagger}e^{\hat a}. \nonumber
\end{equation}

This simple example illustrates the central idea of the present approach: the normal-ordering problem is reduced to determining scalar functions
from a system of ordinary differential equations. We next apply the same procedure to exponentials quadratic in bosonic operators, whose generators form the $su(1,1)$ algebra.

\subsection{The $su(1,1)$ Algebra and Normal Ordering of Exponentials Quadratic in Bosonic Operators }
\label{subsec:2.3}

We now extend the method developed in the previous subsection to exponentials quadratic in bosonic creation and annihilation operators.
Various special cases, particularly squeezing transformations, are well known in the literature~\cite{walls2008, gerry2005, yuen1976, stoler1970, stoler1971} and can be treated using standard factorization techniques \cite{wei1963, wilcox1967, truax1985}. Here, we consider the general
quadratic form and show how the present procedure provides a systematic and transparent derivation of its normal-ordered representation. The
relevant algebraic structure is the $su(1,1)$ algebra generated by quadratic combinations of the creation and annihilation operators.

Indeed, introducing the operators
\begin{equation}\label{eq12}
\hat K_+ = \frac{1}{2}(\hat a^\dagger)^2, \qquad
\hat K_- = \frac{1}{2}\hat a^2, \qquad
\hat K_0 = \frac{1}{2}\left(\hat a^\dagger \hat a + \frac{1}{2}\right),
\end{equation}
and using the canonical commutation relation \( [\hat a,\hat a^\dagger]=1 \),
we readily verify that
\begin{equation}\label{eq13}
[\hat K_0,\hat K_\pm]=\pm \hat K_\pm, \qquad
[\hat K_-,\hat K_+]=2\hat K_0.
\end{equation}
Thus, the operators  \((\hat K_+,\hat K_-,\hat K_0)\)
generate the $su(1,1)$ algebra. 

We therefore consider the general exponential operator
\begin{equation}\label{eq14}
\hat B=e^{\alpha \hat K_++\beta \hat K_-+\gamma \hat K_0},
\end{equation}
where $\alpha$, $\beta$, and $\gamma$ are constants. In analogy with the linear case considered in Subsec.~\ref{subsec:2.2}, we introduce an auxiliary parameter $t$ and define the operator
\begin{equation}\label{eq15}
\hat B(t)=e^{t\left(\alpha\hat K_++\beta \hat K_-+\gamma \hat K_0\right)}=e^{f(t)\hat K_+}e^{g(t)\hat K_0}e^{h(t)\hat K_-},
\end{equation}
where $f(t)$, $g(t)$, and $h(t)$ are scalar functions with the initial conditions
\begin{equation}\label{eq16}
f(0)=g(0)=h(0)=0.
\end{equation}
Since these generators form a closed Lie algebra, the Wei--Norman factorization can be written in the ordered form of Eq.~\eqref{eq15}.

Differentiating Eq.~\eqref{eq15} and then multiplying the resulting equation from the right by $\hat B^{-1}(t)$, which is defined as 
\begin{equation}\label{eq17}
\hat B^{-1}(t)=e^{-t\left(\alpha \hat K_++\beta \hat K_-+\gamma \hat K_0\right)}=e^{-h(t)\hat K_-}e^{-g(t)\hat K_0}e^{-f(t)\hat K_+},
\end{equation}
we arrive at an operator equation
\begin{eqnarray}\label{eq18}
\alpha \hat K_+ + \beta \hat K_- + \gamma \hat K_0
&=f'(t)\,{\hat K_+} +g'(t)\,e^{f(t) {\hat K_+}}{\hat K_0}\, e^{-f (t) {\hat K_+}}\nonumber\\
&\quad+h'(t) \,e^{f(t){\hat K_+}} e^{g(t){\hat K_0}} {\hat K_-}\, e^{-g(t){\hat K_0}} e^{-f(t){\hat K_+}}.
\end{eqnarray}
On the other hand, the BCH relation together with the commutation relations \eqref{eq13} gives formulas
\begin{eqnarray}
e^{f\hat K_+}\hat K_0 e^{-f\hat K_+}&=&\hat K_0-f\hat K_+,\nonumber\\
e^{g\hat K_0}\hat K_- e^{-g\hat K_0}&=&e^{-g}\hat K_-,\nonumber\\
e^{f\hat K_+}\hat K_- e^{-f\hat K_+}&=&\hat K_- -2f\hat K_0 + f^2\hat K_+,\nonumber
\end{eqnarray}
whose substitution into Eq.~\eqref{eq18} yields an operator equation
\begin{eqnarray}\label{eq19}
\alpha\hat K_+ + \beta \hat K_- + \gamma \hat K_0
&=&\Big[f'(t)-f(t)g'(t)+f^2(t)e^{-g(t)}h'(t)\Big]\hat K_+ \nonumber\\
&&\quad+\Big[g'(t)-2f(t)e^{-g(t)}h'(t)\Big]\hat K_0 +e^{-g(t)}h'(t){\hat K_{-}}\,.
\end{eqnarray}
Furthermore, by comparing two sides of Eq.~\eqref{eq19}, we obtain the system of differential equations
\begin{eqnarray}
&&f'(t)-f(t)g'(t)+f^2(t)e^{-g(t)}h'(t)=\alpha, \label{eq20}\\
&&g'(t)-2f(t)e^{-g(t)}h'(t)=\gamma,\label{eq21}\\
&&e^{-g(t)}h'(t)=\beta,\label{eq22}
\end{eqnarray}
which can be solved analytically.

Eliminating \(g'(t)\) and \(h'(t)\) from Eqs.~\eqref{eq20}--\eqref{eq22}, one obtains the Riccati equation
\begin{equation}\label{eq23}
f'(t)-\beta f^2(t)-\gamma f(t)-\alpha=0.
\end{equation}
The form of its solution depends on the sign of $\gamma^2-4\alpha\beta$, leading to hyperbolic or trigonometric
functions. For the case when $\gamma^2-4\alpha\beta>0$, the solution satisfying the initial condition $f(0)=0$ is
\begin{equation}\label{eq24}
f(t)=\frac{2\alpha\,\sinh(\Delta t/2)}
{\Delta\,\cosh(\Delta t/2)-\gamma\, \sinh(\Delta t/2)},
\end{equation}
where $\Delta=\sqrt{\gamma^2-4\alpha\beta}$. 
For the case $\gamma^2-4\alpha\beta<0$, the corresponding expressions are obtained through the replacement $\Delta \rightarrow i\Delta$, yielding trigonometric rather than hyperbolic functions.

Substituting Eq.~\eqref{eq24} into Eq.~\eqref{eq21} and integrating yields
\begin{equation}\label{eq25}
g(t)=-2\ln\left[\cosh\left({\Delta t/2}\right)-\frac{\gamma}{\Delta}
\sinh\left({\Delta t/2}\right)\right],
\end{equation}
satisfying the initial condition $g(0)=0$. Substituting Eq.~\eqref{eq25} into Eq.~\eqref{eq22} yields
\begin{equation}\label{eq26}
h(t)=\frac{2\beta\,\sinh(\Delta t/2)}{\Delta\cosh(\Delta t/2)-\gamma\,\sinh(\Delta t/2)}
\end{equation}
with the initial condition $h(0)=0$.

Setting $t=1$ in Eq.~\eqref{eq15}, we obtain
\begin{eqnarray}\label{eq27}
\hat B=e^{\alpha\,\hat K_++\beta\,\hat K_-+\gamma\,\hat K_0}
=e^{f(1)\hat K_+}e^{g(1)\hat K_0}e^{h(1)\hat K_-}\,.
\end{eqnarray}
A particularly important special case is the squeezing operator
\begin{equation}\label{eq28}
\hat S(\xi)=\exp\left[\frac{\xi}{2} (\hat a^\dagger)^2-\frac{\xi^*}{2}\hat a^2\right],
\end{equation}
which corresponds to $\alpha=\xi,\,\beta=-\xi^*,\,\gamma=0$. In this case, $\Delta=\sqrt{\gamma^2-4\alpha\beta}=2\sqrt{\xi\xi^*}=2r$. Here, we use the notation $\xi=re^{i\phi}$. Substituting the corresponding functions into Eq.~\eqref{eq27} gives
\begin{equation}\label{eq29}
\hat S(\xi)=e^{\eta \hat K_+}e^{\ln(1-|\eta|^2)\hat K_0}e^{-\eta^*\hat K_-},
\end{equation}
where $\eta=e^{i\phi}\tanh r$.

The squeezing operator thus provides an important special case of the general $su(1,1)$ factorization derived above. Together with the
Heisenberg--Weyl example of Subsec.~\ref{subsec:2.2}, this demonstrates how the same algebraic procedure constructs normal-ordered representations for exponential operators associated with different closed bosonic Lie algebras. In the next section, we extend this approach to two-mode bosonic systems and subsequently apply it to the analytical evaluation of excitonic form factors.

\section{Two-Mode Bosonic Operators and Excitonic Form Factors }
\label{sec:3}

In the preceding section, we developed the normal-ordering procedure for one-mode bosonic operators associated with the Heisenberg--Weyl and $su(1,1)$ algebras. We now extend this approach to two-mode bosonic systems, for which different closed algebraic structures emerge depending on the combinations of creation and annihilation operators involved~\cite{schwinger1965, perelomov1986}. In particular, number-conserving combinations such as $\hat a^\dagger\hat b$ and $\hat b^\dagger\hat a$ generate the $su(2)$ algebra, whereas pair-creation and pair-annihilation combinations such as $\hat a^\dagger\hat b^\dagger$ and $\hat a\hat b$ generate the $su(1,1)$ algebra.

These two-mode algebraic structures occur widely in quantum physics. The $su(2)$ realization describes, for example, mode mixing, beam-splitter transformations, interferometry, polarization optics, and coupled-mode dynamics~\cite{yurke1986, walls2008, gerry2005}, while the $su(1,1)$ realization underlies two-mode squeezing, parametric amplification, Bogoliubov transformations, and pair-creation processes~\cite{yuen1976, stoler1970, stoler1971, truax1985, yurke1986}. In the present work, our main interest in these algebras arises from the analytical evaluation of excitonic form factors. After illustrating the normal-ordering procedure for two-mode exponential operators associated with the $su(2)$ algebra, we formulate the excitonic form-factor problem in the bosonic representation and apply the algebraic method to the exponential operators generated by the Laplace and Fourier representations of the Rytova--Keldysh potential.

\subsection{Normal Ordering of Two-Mode Exponential Operators Associated with the $su(2)$ Algebra }
\label{subsec:3.1}

We consider two independent bosonic modes described by annihilation and creation operators $(\hat a,\hat a^\dagger)$
and $(\hat b,\hat b^\dagger)$, which satisfy the commutation relations
\begin{equation}\label{eq30}
[\hat a,\hat a^\dagger]=1, \qquad [\hat b,\hat b^\dagger]=1,
\end{equation}
while all remaining commutators vanish.
Introducing the operators
\begin{equation}\label{eq31}
\hat J_+ = \hat a^\dagger \hat b, \qquad \hat J_- = \hat b^\dagger \hat a, 
\qquad \hat J_0 = \frac{1}{2} \left(\hat a^\dagger \hat a-\hat b^\dagger \hat b\right),
\end{equation}
and using Eq.~\eqref{eq30}, we readily verify that
\begin{equation}\label{eq32}
[\hat J_0,\hat J_\pm]=\pm \hat J_\pm, \qquad [\hat J_-,\hat J_+]=-2\hat J_0.
\end{equation}
Therefore, the operators $(\hat J_+,\hat J_-,\hat J_0)$ generate the $su(2)$ algebra.

We therefore consider the general exponential operator
\begin{equation}\label{eq33}
\hat U=e^{\alpha\,\hat J_+ + \beta\,\hat J_- + \gamma\,\hat J_0},
\end{equation}
where $\alpha$, $\beta$, and $\gamma$ are constants. Following the procedure developed in Subsec.~\ref{subsec:2.3}, we introduce an auxiliary parameter $t$ and define
\begin{equation}\label{eq34}
\hat U(t)=e^{t(\alpha\,\hat J_+ + \beta\,\hat J_- + \gamma\,\hat J_0)}
=e^{f(t)\hat J_+}e^{g(t)\hat J_0}e^{h(t)\hat J_-},
\end{equation}
where $f(t)$, $g(t)$, and $h(t)$ are scalar functions satisfying the initial conditions
\begin{equation}\label{eq35}
f(0)=g(0)=h(0)=0.
\end{equation}
The functions $f(t)$, $g(t)$, and $h(t)$ are determined below, after which we set $t=1$ to obtain the factorized form of $\hat U$. 

Using the $su(2)$ commutation relations, we obtain the system
\begin{equation}
f'(t)+\beta f^2(t)-\gamma f(t)-\alpha=0,
\label{eq36}
\end{equation}
together with
\begin{equation}
g'(t)=\gamma-2\beta f(t),
\qquad
h'(t)=\beta e^{g(t)}.
\label{eq37}
\end{equation}
The sign difference between Eq.~\eqref{eq36} and the corresponding $su(1,1)$ equation, Eq.~\eqref{eq23}, originates from the different commutation relations of the two algebras.

For these equations, the solutions satisfying the initial condition \eqref{eq35} have the same analytical structure as in the $su(1,1)$ case. For $\gamma^2+4\alpha \beta<0$, we obtain
\begin{eqnarray}\label{eq38}
f(t)&=&\frac{2\alpha\,\sin(\Delta t/2)}{\Delta\cos(\Delta t/2)-\gamma\,\sin(\Delta t/2)},\nonumber\\
g(t)&=&-2\ln\left[\cos\left({\Delta t}/{2}\right) -\frac{\gamma}{\Delta} \sin(\Delta t/2)\right],\\
h(t)&=&\frac{2\beta\sin(\Delta t/2)}{\Delta\cos(\Delta t/2)-\gamma\sin(\Delta t/2)},\nonumber
\end{eqnarray}
where $\Delta=\sqrt{\abs{\gamma^2+4\alpha\beta}}$. In the case $\gamma^2+4\alpha \beta>0$, the corresponding expressions involve hyperbolic functions, analogous to Eq.~\eqref{eq24}, rather than the trigonometric form in Eq.~\eqref{eq38}.

An important special case corresponds to the beam-splitter transformation
\begin{equation}\label{eq39}
\hat U_{\mathrm{BS}}=\exp\left[\theta\left(e^{i\phi}\hat a^\dagger \hat b
-e^{-i\phi}\hat b^\dagger \hat a\right)\right],
\end{equation}
which is obtained by choosing
$\alpha=\theta e^{i\phi}$, $ \beta=-\theta e^{-i\phi}$, $\gamma=0$. 
For this choice, $\Delta=2\theta$, yielding
\begin{eqnarray} \label{eq40}
f(1)=e^{i\phi} \tan \theta,
\quad g(1)=-2\ln\left(\cos {\theta}\right),\quad h(1)=-e^{-i\phi} \tan \theta. 
\end{eqnarray}
The factorized normal-ordered form of $\hat U_{BS}$ is obtained directly from Eqs.~\eqref{eq34} and \eqref{eq40} as
\begin{equation}\label{eq41}
\hat U_{BS}
=e^{e^{i\phi} \tan \theta\, {\hat a^\dagger \hat b}}
e^{-\ln\left(\cos {\theta}\right) {(\hat a^\dagger \hat a - \hat b^\dagger \hat b)}}e^{-e^{-i\phi} \tan \theta\, {\hat b^\dagger \hat a}}.
\end{equation}

The above result illustrates that the same algebraic procedure developed for one-mode systems extends naturally to two-mode exponential operators. We now turn to its application to the analytical evaluation of excitonic form factors.

\subsection{Bosonic Formulation of Excitonic Form Factors}
\label{subsec:3.2}

The normal-ordering procedure developed above is particularly useful for the analytical evaluation of matrix elements between oscillator
states. An important application arises in excitonic problems in two-dimensional semiconducting materials, where matrix elements of the
Rytova--Keldysh interaction can be expressed in terms of exponential form factors. Depending on the symmetry of the system, the Laplace or
Fourier representation of the interaction provides the appropriate starting point for their analytical evaluation.

In coordinate space, the Rytova--Keldysh potential is most commonly written via the Struve function and the Bessel function of the second kind, which is suitable for numerical calculations~\cite{berkelbach2013}. However, for analytical treatments based on algebraic methods, as used in several recent studies \cite{ly2023, ly2025}, the Rytova--Keldysh potential can be written in the equivalent Laplace-type form
 \begin{equation}\label{eq42}
V(r)=-\frac{e^2}{\kappa r_0}   \int\limits_0^{+\infty} \frac{e^{-(r/r_0)\,t}}{\sqrt{1+t^2}}dt
=-\frac{e^2}{\kappa}   \int\limits_0^{+\infty}\frac{e^{-r\,t}}{\sqrt{1+r_0^2 \, t^2}}dt,
\end{equation}
where $r_0$ and $\kappa$ are the screening length and effective dielectric constant, respectively.
Evaluation of matrix elements of the Rytova--Keldysh potential, therefore, reduces to calculating the form factor
$\langle e^{-rt}\rangle$.

Recently, the algebraic approach has been applied to the exciton problem in anisotropic two-dimensional materials \cite{le2026prb,le2026cip}, in which the space scaling yields the anisotropic Rytova--Keldysh potential $V(x, y/\beta)$ with the parameter $\beta$ characterizing the space anisotropy. 
In this case, the Rytova-Keldysh potential is more conveniently represented in Fourier form,
\begin{equation}\label{eq43}
V(x, y/\beta)=-\frac{e^2}{2\pi \kappa }   \int\limits_0^{+\infty} \frac{e^{i (q_x x+ q_y y/\beta) }}{ q(1+r_0 q)}d^2 q
=-\frac{e^2\beta}{2\pi \kappa }   \int\limits_0^{+\infty} \frac{e^{i \bf{q}\cdot \bf{r} }}{ q_{\beta}(1+r_0 q_{\beta})} d^2 q,
\end{equation}
where $q=\sqrt{q_x^2+q_y^2}$ and $q_{\beta}=\sqrt{q_x^2+\beta^2 q_y^2}$. 
Therefore, the evaluation of matrix elements of the anisotropic Rytova--Keldysh potential reduces to the calculation of the form factor
 $\langle e^{i {\bf q} \cdot {\bf r}} \rangle$.

Consequently, the calculation of exciton energies in two-dimensional semiconducting materials requires the evaluation of the form factors
$\langle e^{-rt}\rangle$ or $\langle e^{i{\bf q}\cdot{\bf r}}\rangle$, depending on whether the system is isotropic or anisotropic. As demonstrated in our previous studies~\cite{nguyen2019, ly2023, ly2025, le2026cip, le2026prb}, the Levi--Civita transformation
\begin{equation}\label{eq44}
x=u^2-v^2,\qquad y=2uv
\end{equation}
maps the excitonic problem onto an anharmonic oscillator problem and provides a convenient framework for its algebraic treatment. The corresponding bosonic representation is introduced through the annihilation and creation operators
\begin{eqnarray}\label{eq45} 
&\hat{a} = {\dfrac{1}{\sqrt{2}}}\left(\xi+\dfrac{\partial}{\partial{\xi}^*}\right), 
\quad \hat{a}^\dagger = {\dfrac{1}{\sqrt{2}}}\left(\xi^*-\dfrac{\partial}{\partial{\xi}}\right), \nonumber\\ 
&\hat{b} = {\dfrac{1}{\sqrt{2}}}\left(\xi^*+\dfrac{\partial}{\partial{\xi}}\right), 
\quad \hat{b}^\dagger = {\dfrac{1}{\sqrt{2}}}\left(\xi-\dfrac{\partial}{\partial{\xi}^*}\right), 
\end{eqnarray} 
where the complex coordinates are defined by $ \xi=u+iv $ and $ \xi^*=u-iv. $ These operators satisfy the bosonic commutation relations \eqref{eq30}.

In this representation, the distance \(r\) and the Cartesian coordinates \(x\) and \(y\) are expressed in terms of the bosonic operators as \begin{eqnarray}\label{eq46} 
r&=&\frac{1}{2} \left(\hat{a}^\dagger\hat{a}+\hat{b}^\dagger\hat{b}+1+\hat{a}\hat{b}+\hat{a}^\dagger\hat{b}^\dagger\right),\nonumber\\ x&=&\frac{1}{4}\left[{\hat{a}}^2+ {\hat{b}}^2 +({\hat{a}^\dagger})^2 
+ ({\hat{b}^\dagger})^2+2\hat{b}^\dagger\hat{a}+2\hat{a}^\dagger\hat{b}\right], \\
 y&=&\frac{i}{4}\left[-{\hat{a}}^2+ {\hat{b}}^2 +({\hat{a}^\dagger})^2 - ({\hat{b}^\dagger})^2-2\hat{b}^\dagger\hat{a}
+2\hat{a}^\dagger\hat{b}\right]. \nonumber
\end{eqnarray} 
The corresponding oscillator basis is 
\begin{equation}\label{eq47} 
 |n,m\rangle =\frac{1}{\sqrt{(n+m)!(n-m)!}}
         (\hat a^\dagger)^{n+m} (\hat b^\dagger)^{n-m} |0\rangle
\end{equation} 
where \( n=0,1,2,\ldots \) and \( m=-n,-n+1,\ldots,n. \) The vacuum state satisfies 
\begin{equation}\label{eq48} 
\hat a|0\rangle=0, \qquad \hat b|0\rangle=0. 
\end{equation}

With these relations, the relevant excitonic form factors are reduced to matrix elements of exponential operators expressed entirely in terms
of bosonic creation and annihilation operators. In the following subsections, we derive their normal-ordered representations and evaluate
the corresponding matrix elements for isotropic and anisotropic excitons.

\subsection{Isotropic Exciton Form Factor  $\langle n, m| e^{-rt}|j,m'\rangle$}
\label{subsec:3.3}

We first consider the form factor arising from the Laplace representation of the Rytova--Keldysh
potential for isotropic excitons. 

Using the bosonic representation of $r$ in Eq.~\eqref{eq46}, we introduce the operators
\begin{equation}\label{eq49}
\hat M= {\hat a} {\hat b},\qquad  \hat M^+= {\hat a^\dagger} {\hat b^\dagger}, 
\qquad \hat N=\frac{1}{2}\left({\hat a^\dagger}{\hat a}+{\hat b^\dagger}{\hat b}+1\right),
\end{equation}
which satisfy the commutation relations:
\begin{equation}\label{eq50}
\left[\hat M, \hat M^+\right]= 2\hat N, \quad \left[\hat N, \hat M^+\right]= \hat M^+, 
\quad \left[\hat N, \hat M\right]= -\hat M.
\end{equation}
Thus, the operators $\hat M$, $\hat M^+$, and $\hat N$ generate the $su(1,1)$ algebra. Since the operator $r$ can be
expressed entirely in terms of these generators, the exponential $e^{-rt}$ can be factorized using the procedure developed in Subsec.~\ref{subsec:2.3} as
\begin{equation}\label{eq51}
e^{-rt}=e^{-\frac{t}{2}\left(\hat M + \hat M^+ +2\hat N\right)}=e^{f(t) \hat M^+}e^{g(t) \hat N} e^{h(t) \hat M}
\end{equation}

For the present case, we have $\alpha=-\frac12, \beta=-\frac12, \gamma=-1,$ which yields
$\Delta=\sqrt{\gamma^2-4\alpha\beta}=0.$
Since Eq.~\eqref{eq24} takes an indeterminate form in the limit $\Delta\rightarrow0$, we solve the Riccati equation~\eqref{eq23}
directly, which in the present case becomes
\begin{equation}\label{eq52}
f'(t)+\frac{1}{2}\left[f(t)+1\right]^2=0.
\end{equation}
Solving Eq.~\eqref{eq52} together with Eqs.~\eqref{eq21} and \eqref{eq22} yields
\begin{equation}\label{eq53}
f(t)=-\frac{t}{t+2}, \qquad
g(t)=-2\ln\left(1+\frac{t}{2}\right), \qquad
h(t)=-\frac{t}{t+2},
\end{equation}
which satisfy the initial conditions \(f(0)=g(0)=h(0)=0.\)

The action of the operators $\hat M$, $\hat M^+$, and $\hat N$ on the basis states is
\begin{eqnarray}\label{eq54}
{\hat M}^j\, |n,m\rangle&=&\sqrt{\frac{(n-m)!(n+m)!}{(n-m-j)!(n+m-j)!}}\,|n-j,m\rangle,\quad (j=1,2,...,n-\abs{m})\nonumber\\
({\hat M}^+)^j\, |n,m\rangle&=&\sqrt{\frac{(n-m+j)!(n+m+j)!}{(n-m)!(n+m)!}}\,|n+j,m\rangle,\\
{\hat N}^j \,|n,m\rangle &=&\left(n+\frac{1}{2}\right)^j\,|n,m\rangle,\nonumber
\end{eqnarray}
which follow directly from Eqs.~\eqref{eq47} and \eqref{eq48} together with the commutation relations~\eqref{eq30}. 
Using the normal-ordered representation~\eqref{eq51}, the form factor can be written as
\begin{eqnarray}\label{eq55}
F_{nm;jm'}&=&\langle n, m| e^{-tr} |j, m'\rangle =\langle n, m| e^{f(t) \hat M^+}e^{g(t) \hat N}e^{h(t) \hat M} 
|j, m'\rangle\nonumber\\
&=&\sum\limits_{k_1,m_1}\sum\limits_{k_2,m_2}\langle n, m| e^{f(t) \hat M^+}|k_1,m_1\rangle 
\langle k_1,m_1| e^{g(t) \hat N} |k_2,m_2\rangle 
\langle k_2,m_2|e^{h(t) \hat M}|j, m'\rangle\,.
\end{eqnarray}
Using the orthogonality $\langle n,m |j,m'\rangle=\delta_{nj}\delta_{mm'}$ and formulas \eqref{eq54}, we can calculate all elements in the sum \eqref{eq55} as
\begin{eqnarray}\label{eq56}
\langle k_1,m_1| e^{g(t) \hat N} |k_2,m_2\rangle &=&e^{g(t) \left(k_2+\frac{1}{2}\right)}\delta_{k_1k_2}\delta_{m_1m_2}
=\left( \frac{2}{t+2}\right)^{2k_2+1}\delta_{k_1k_2}\delta_{m_1m_2}\,,\\
\label{eq57}
\langle n,m| e^{f(t) \hat M^+} |k_1,m_1\rangle &=&\sum\limits_{l=0}^{+\infty}
\frac{f^l}{l!}\langle n,m| ({\hat M^+})^l |k_1,m_1\rangle \nonumber\\
&=&\sum\limits_{l=0}^{+\infty}\frac{1}{l!}\left( \frac{-t}{t+2}\right)^l
\sqrt{\frac{(k_1-m_1+l)!}{(k_1-m_1)!}}\sqrt{\frac{(k_1+m_1+l)!}{(k_1+m_1)!}}
\delta_{n,k_1+l}\delta_{mm_1}\nonumber\\
&=&\sqrt{\binom{n-m}{n-k_1}\binom{n+m}{n-k_1}}\left( \frac{-t}{t+2}\right)^{n-k_1}\,\delta_{mm_1}\,,
\end{eqnarray}
\begin{eqnarray}\label{eq58}
\langle k_2,m_2| e^{h(t) \hat M} |j,m'\rangle &=&\sum\limits_{l=0}^{+\infty}
\frac{h^l}{l!}\langle k_2,m_2| {\hat M}^l |j,m'\rangle \nonumber\\
&=&\sum\limits_{l=0}^{+\infty}\frac{1}{l!}\left( \frac{-t}{t+2}\right)^l
\sqrt{\frac{(j-m')!}{(j-m'-l)!}}\sqrt{\frac{(j+m')!}{(j+m'-l)!}}
\delta_{k_2,j-l}\delta_{m_2m'}\nonumber\\
&=&\left( \frac{-t}{t+2}\right)^{j-k_2}
\sqrt{\binom{j-m'}{j-k_2}\binom{j+m'}{j-k_2}}\,\delta_{m_2m'}\,.
\end{eqnarray}
Substituting Eqs.~\eqref{eq56}, \eqref{eq57}, and \eqref{eq58} into Eq.~\eqref{eq55}, we obtain the form factor as
\begin{eqnarray}\label{eq59}
F_{nm;jm'}(t)
=(-1)^{n+j}\delta_{mm'}\sum\limits_{k=0}^{\min(n,j)}
\sqrt{\binom{n-m}{n-k}\binom{n+m}{n-k}{\binom{j-m'}{j-k}\binom{j+m'}{j-k}}}\, 
 \,\frac{ \left(\dfrac{t}{2}\right)^{n+j-2k} }  {\left(1+\dfrac{t}{2}\right)^{n+j+1}}.
\end{eqnarray}

Equation~\eqref{eq59} provides an explicit analytical expression for the isotropic exciton form factor. In our previous work based on the Laplace representation of the Rytova--Keldysh potential~\cite{ly2023}, this form factor was not presented separately; instead, the calculation was carried through directly to the final matrix elements of the interaction. Here, we derive the form factor explicitly from the general $su(1,1)$ factorization developed in Subsec.~\ref{subsec:2.3}, thereby making the underlying normal-ordering procedure and algebraic structure transparent. Since Eq.~\eqref{eq59} is obtained independently of the specific form of the Rytova--Keldysh interaction, it may also be useful in other problems involving radial exponential operators or interactions admitting a Laplace-type representation.

We now turn to the anisotropic case, where the Fourier representation leads to a more complicated exponential operator and requires a larger
closed algebra for the evaluation of the corresponding form factor.

\subsection{Anisotropic Exciton Form Factor  $\langle n, m| e^{i {\bf q}\cdot {\bf r}}|j,m'\rangle$}
\label{subsec:3.4}

Using the bosonic representations of $x$ and $y$ in Eq.~\eqref{eq46}, we first express the exponent $i{\bf q}\cdot{\bf r}$ in terms of the
creation and annihilation operators as
\begin{eqnarray}\label{eq60}
i{\bf q}\cdot {\bf r}=iq_1 x+ iq_2 y&=&\frac{1}{4}\left( iq_1+q_2\right){\hat a}^2+\frac{1}{4}\left( iq_1-q_2\right){\hat b}^2
+\frac{1}{4}\left( iq_1- q_2\right){(\hat a^\dagger)}^2\nonumber\\
&&+\frac{1}{4}\left( iq_1+q_2\right){(\hat b^\dagger)}^2+\frac{1}{2}\left( iq_1+q_2\right){\hat b^\dagger \hat a}
+\frac{1}{2}\left( i  q_1-q_2\right){\hat a^\dagger \hat b}\,.
\end{eqnarray}
The operator structure in Eq.~\eqref{eq60} suggests introducing the following six generators:
\begin{eqnarray}\label{eq61}
\hat A&=&\frac{iq_1+q_2}{2\sqrt{{q_1}^2+{q_2}^2}}{\hat a}^2+\frac{iq_1-q_2}{2\sqrt{{q_1}^2+{q_2}^2}}{\hat b}^2\,,\nonumber\\
\hat A^+&=&\frac{-iq_1+q_2}{2\sqrt{{q_1}^2+{q_2}^2}}{(\hat a^\dagger)}^2-\frac{iq_1+q_2}{2\sqrt{{q_1}^2+{q_2}^2}}{(\hat b^\dagger)}^2\,,\nonumber\\
\hat K&=&\frac{iq_1+q_2}{2\sqrt{{q_1}^2+{q_2}^2}}{\hat b^\dagger \hat a}+
\frac{iq_1-q_2}{2\sqrt{{q_1}^2+{q_2}^2}}{\hat a^\dagger \hat b}\,,\\
\hat M&=& \hat a \hat b\,,\quad \hat M^+=\hat a^\dagger \hat b^\dagger\,,\quad \hat N =\frac{1}{2}\left( \hat a^\dagger \hat a + \hat b^\dagger \hat b +1  \right)\,,\nonumber
\end{eqnarray}
which satisfy the commutation relations:
\begin{eqnarray}\label{eq62}
&\left[\hat N, \hat A^+  \right]= \hat A^+\,,\quad \left[\hat N, \hat A  \right]= - \hat A\,,\quad
\left[\hat A, \hat A^+  \right]=2 \hat N\,,\nonumber\\
&\left[\hat K, \hat A \right]=\hat M\,,\quad\qquad \left[\hat K, \hat A^+  \right]=\hat M^+\,,\nonumber\\
&\left[\hat K, \hat M  \right]=-\hat A\,,\quad\qquad \left[\hat K, \hat M^+  \right]=-\hat A^+\,, \nonumber\\
&\left[\hat A, \hat M^+  \right]=2\hat K\,,\quad\qquad \left[\hat A^+, \hat M  \right]=2\hat K\,,\\
&\left[\hat N, \hat M^+  \right]=\hat M^+\,,\quad \left[\hat N, \hat M  \right]=-\hat M\,,\quad
 \left[\hat M, \hat M^+  \right]=2 \hat N\,, \nonumber\\
&\left[\hat A, \hat M  \right]=0\,,\quad \left[\hat A^+, \hat M^+  \right]=0\,,\quad \left[\hat N, \hat K  \right]=0\,.\nonumber
\end{eqnarray}
The commutation relations~\eqref{eq62} show that the six operators $\hat A$, $\hat A^+$, $\hat M$, $\hat M^+$, $\hat N$, and $\hat K$ form a closed Lie algebra. By appropriate linear combinations of the generators, this algebra can be decomposed into two mutually commuting $su(1,1)$ algebras and is therefore isomorphic to $su(1,1)\oplus su(1,1)\simeq so(2,2)$~\cite{gerry2000}.
In the present basis, the algebra contains two $su(1,1)$ subalgebras generated by $(\hat A,\hat A^+,\hat N)$ and $(\hat M,\hat M^+,\hat N)$, respectively, which share the generator $\hat N$ and are coupled through $\hat K$. The remaining commutation relations in Eq.~\eqref{eq62} describe the mixing between these sectors. The closure of this six-generator algebra allows the Wei--Norman factorization procedure to be applied to the exponential operator
\begin{equation}\label{eq63}
e^{i{\bf q}\cdot{\bf r}}
=e^{\frac{q}{2}\left(\hat A-\hat A^++2\hat K\right)},
\end{equation}
where $q=\sqrt{q_1^2+q_2^2}$.

Following the algebraic procedure developed in the preceding sections, we introduce an auxiliary parameter $t$ and write
\begin{equation}\label{eq64}
e^{t\left(\hat A-\hat A^++2\hat K\right)}
=e^{f(t)\hat A^+}e^{g(t)\hat M^+}
 e^{h(t)\hat K}e^{o(t)\hat N}
 e^{p(t)\hat M}e^{\ell(t)\hat A},
\end{equation}
with $t=q/2$. The ordering in Eq.~\eqref{eq64} is chosen to provide a normal-ordered representation with respect to the quadratic bosonic generators: the creation-type generators $\hat A^+$ and $\hat M^+$ appear on the left, the annihilation-type generators $\hat M$ and $\hat A$ on the right, while $\hat K$ and $\hat N$ are neutral generators and are placed between them. The scalar functions in Eq.~\eqref{eq64} satisfy the initial conditions
\begin{equation}\label{eq65}
f(0)=g(0)=h(0)=o(0)=p(0)=\ell(0)=0\,.
\end{equation}

Differentiating Eq.~\eqref{eq64}, multiplying the resulting equation from the right by the inverse operator, and using the BCH relations together with the commutation relations~\eqref{eq62}, we obtain a closed system of differential equations for the six scalar functions:
\begin{eqnarray}\label{eq66}
f'(t) &=& f^2(t) - \left[1+g(t)\right]^2,\qquad\,
g'(t) =2f(t)\left[1+g(t)\right]\,,\nonumber\\
h'(t) &=& 2\left[1+g(t)\right]\,,\qquad\qquad\quad
o'(t) = 2f(t)\,,\\
p'(t) &=& -e^{o(t)}\sin\left(h(t)\right)\,,\qquad\quad\,\,
\ell'(t) = e^{o(t)}\cos\left(h(t)\right)\,.\nonumber
\end{eqnarray}
Solving equations \eqref{eq66} with the initial conditions \eqref{eq65}, we obtain
\begin{eqnarray}\label{eq67}
f(t)&=&-\frac{t}{1+t^2}\,,\qquad\, g(t)=-\frac{t^2}{1+t^2}\,,\nonumber\\
h(t)&=&2 \arctan {t}\,, \qquad o(t)=-\ln{(1+t^2)}\,,\\
\ell(t)&=&\frac{t}{1+t^2}\,,\qquad\quad p(t)=-\frac{t^2}{1+t^2}\,.\nonumber
\end{eqnarray}
Substituting the solutions~\eqref{eq67} into Eq.~\eqref{eq64} with $t=q/2$, we obtain
\begin{eqnarray}\label{eq68}
e^{i{\bf q}\cdot{\bf r}}&=&e^{q_- {(\hat a^\dagger)^2}} e^{q_+ {(\hat b^\dagger)^2}}e^{-q_0 \hat a^\dagger \hat b^\dagger}
e^{2\arctan (q/2) \hat K} \nonumber\\
&&\times e^{-(\hat a^\dagger \hat a + \hat b^\dagger \hat b +1)\ln{(\sqrt{q^2+4}/2)}}
e^{-q_0 \hat a \hat b} e^{q_+ {\hat a^2}} e^{q_- {\hat b^2}}\,,
\end{eqnarray}
where
\begin{equation}
q_+=\frac{iq_1+q_2}{q^2+4}\,,\qquad q_-=\frac{iq_1-q_2}{q^2+4}\,,\qquad q_0=\frac{q^2}{q^2+4}\,.\nonumber
\end{equation}

When expressed in terms of the elementary bosonic operators, however, Eq.~\eqref{eq68} is not yet completely normal ordered because the factor $\exp[2\arctan(q/2)\hat K]$ contains both $\hat b^\dagger\hat a$ and $\hat a^\dagger\hat b$. To complete the normal ordering, we exploit the $su(2)$ structure of this remaining operator.
Let us define the following operators 
\begin{eqnarray}\label{eq69}
\hat K_-&=&\frac{iq_1+q_2}{\sqrt{{q_1}^2+{q_2}^2}}{\hat b^\dagger \hat a}\,,
\quad \hat K_+=\frac{-iq_1+q_2}{\sqrt{{q_1}^2+{q_2}^2}}{\hat a^\dagger \hat b}\,,
\quad \hat n=\frac{1}{2}\left(\hat a^\dagger \hat a -\hat b^\dagger \hat b\right)\,,
\end{eqnarray}
which satisfy the commutation relations
\begin{eqnarray}\label{eq70}
\left[\hat n, \hat K_\pm  \right]= \pm \hat K_\pm,\quad \left[\hat K_-, \hat K_+  \right]=- 2 \hat n\,.
\end{eqnarray}
Thus, $(\hat K_+,\hat K_-,\hat n)$ generate the $su(2)$ algebra, and the factorization derived in Subsec.~\ref{subsec:3.1} can be applied
directly. Since $\hat K={(\hat K_--\hat K_+)}/{2},$ the present case corresponds to  $\alpha=-\beta=-\arctan{(q/2)}$, $\gamma=0$. 
Since $\gamma^2+4\alpha\beta<0$ in the present case, the trigonometric solution in Eq.~\eqref{eq38} applies, with  $\Delta = \sqrt{|\gamma^2+4\alpha\beta|}=2\arctan{(q/2)}$. As a result, we have 
\begin{eqnarray}\label{eq71}
f(1)=-\frac{1}{2} q\,,\qquad g(1)=\ln{(q^2/4+1)}\,,\qquad h(1)=\frac{1}{2} q\,.
\end{eqnarray}
Plugging functions \eqref{eq71} into \eqref{eq34}, we obtain
\begin{eqnarray}\label{eq72}
e^{2\arctan (q/2) \hat K}=e^{\frac{iq_1-q_2}{2} {\hat a^\dagger \hat b}} e^{\ln{(\sqrt{q^2+4}/2)}(\hat a^\dagger \hat a -\hat b^\dagger \hat b) }e^{\frac{iq_1+q_2}{2} {\hat b^\dagger \hat a}} .
\end{eqnarray}

Combining Eqs.~\eqref{eq68} and \eqref{eq72}, we can now evaluate the anisotropic exciton form factor in the fully normal-ordered
representation:
\begin{eqnarray}\label{eq73}
G_{nm;jm'}&=&\langle n, m| e^{i {\bf q}\cdot{\bf r}} |j, m'\rangle =\langle n, m| e^{q_- {(\hat a^\dagger)^2}} 
e^{q_+ {(\hat b^\dagger)^2}}e^{-q_0 \hat a^\dagger \hat b^\dagger} e^{2\arctan (q/2) \hat K}\nonumber\\
&& \qquad\qquad\times e^{-(\hat a^\dagger \hat a + \hat b^\dagger \hat b +1)\ln{(\sqrt{q^2+4}/2)}}
e^{-q_0 \hat a \hat b} e^{q_+ {\hat a^2}} e^{q_- {\hat b^2}} |j, m'\rangle\nonumber\\
&=&\sum\limits_{k_1,m_1}\sum\limits_{k_2,m_2}
\langle n, m| e^{q_- {(\hat a^\dagger)^2}} e^{q_+ {(\hat b^\dagger)^2}}e^{-q_0 \hat a^\dagger \hat b^\dagger}|k_1,m_1\rangle \\
&&\qquad\times \langle k_1,m_1| e^{2\arctan (q/2) \hat K} e^{-(\hat a^\dagger \hat a + \hat b^\dagger \hat b +1)\ln{(\sqrt{q^2+4}/2)}}
|k_2,m_2\rangle \nonumber\\
&&\quad\qquad\times\langle k_2,m_2|e^{-q_0 \hat a \hat b} e^{q_+ {\hat a^2}} e^{q_- {\hat b^2}}|j, m'\rangle\,.\nonumber
\end{eqnarray}
As in the isotropic calculation leading to Eq.~\eqref{eq55}, we insert complete sets of oscillator states and evaluate separately the three
matrix elements appearing in Eq.~\eqref{eq73}. We denote these contributions by $S_1$, $S_2$, and $S_3$.

Starting from the expansion of the exponential operator in the first matrix element $S_1$, we obtain
\begin{equation}\label{eq74}
\begin{aligned}
S_1&=\langle n, m| e^{q_- {(\hat a^\dagger)^2}} e^{q_+ {(\hat b^\dagger)^2}}
e^{-q_0 \hat a^\dagger \hat b^\dagger}|k_1,m_1\rangle\\
&=\sum_{r,s,l=0}^{\infty}\frac{q_-^{r}}{r!}\frac{q_+^{s}}{s!}\frac{(-q_0)^l}{l!}
\left\langle n,m\left|(\hat a^\dagger)^{2r+l}(\hat b^\dagger)^{2s+l}\right|k_1,m_1\right\rangle .
\end{aligned}
\end{equation}
Using the action of the creation operators on the oscillator basis and the orthogonality of the basis states, one of the summations can be
performed explicitly. After rearranging the remaining summation indices, we obtain
\begin{equation}\label{eq75}
\begin{aligned}
S_1&=\sum_{s=0}^{n-k_1}\sum_{r=0}^{s} \frac{q_-^r q_+^{s-r} (-q_0)^{n-k_1-s}}{r!(s-r)!(n-k_1-s)!}
 \sqrt{\frac{(n+m)!(n-m)!}{(k_1+m_1)!(k_1-m_1)!}}\delta_{m_1,m+s-2r}.
\end{aligned}
\end{equation}

The second matrix element, which contains the $su(2)$ mixing transformation derived above, is evaluated similarly and gives
\begin{equation} \label{eq76}
\begin{aligned} 
S_2 &= \left\langle k_1,m_1\left| e^{c_-\hat a^\dagger \hat b} e^{\lambda(\hat a^\dagger\hat a-\hat b^\dagger\hat b)} e^{c_+\hat b^\dagger \hat a} e^{-\lambda(\hat a^\dagger\hat a+\hat b^\dagger\hat b+1)} \right|k_2,m_2\right\rangle \\
&= \delta_{k_1k_2} \sum_{l=0}^{\min(k_2+m_1,k_2+m_2)} c_+^{\,k_2+m_2-l} c_-^{\,k_2+m_1-l} 
 \left(\frac{\sqrt{q^2+4}}{2}\right)^{2l-4k_2-1} \\ 
&\quad\times \sqrt{ \binom{2k_2-l}{k_2+m_2-l} \binom{2k_2-l}{k_2+m_1-l} \binom{k_2+m_2}{l} \binom{k_2+m_1}{l} }. 
\end{aligned} 
\end{equation} 
where 
\begin{equation} 
c_+=\frac{iq_1+q_2}{2},\qquad c_-=\frac{iq_1-q_2}{2},\qquad \lambda=\ln\left(\frac{\sqrt{q^2+4}}{2}\right). \nonumber
\end{equation} 

The third matrix element is evaluated in the same manner, yielding
\begin{equation} \label{eq77}
\begin{aligned}
S_3&= \left\langle k_2,m_2\left| e^{-q_0\hat a\hat b} e^{q_+\hat a^2} e^{q_-\hat b^2} \right|j,m'\right\rangle \\
&= \sum_{u=0}^{j-k_2} \sum_{v=0}^{u} \frac{q_+^{\,v} q_-^{\,u-v} (-q_0)^{\,j-k_2-u}}{v!(u-v)!(j-k_2-u)!} 
 \sqrt{ \frac{(j+m')!(j-m')!}{(k_2+m_2)!(k_2-m_2)!} }  \delta_{m_2,m'+u-2v}. 
\end{aligned}
\end{equation} 

Plugging Eqs.~\eqref{eq75}, \eqref{eq76}, and \eqref{eq77} into Eq.~\eqref{eq73}, we obtain
\begin{equation}\label{eq78}
\begin{aligned}
G_{nm;jm'}(q,\theta)&={(-1)^{\frac{m-m'}{2}}}e^{i(m'-m)\theta}\sum_{k=0}^{\min(n,j)}\sum_{s=0}^{n-k}\sum_{r=0}^{s}\sum_{u=0}^{j-k}\sum_{v=0}^{u}\sum_{l=0}^{l_{\max}}
\,\dfrac{\sqrt{\binom{n+m}{k+m}\binom{n-m}{k-m}\binom{j+m'}{k+m'}\binom{j-m'}{k-m'}}}
{\sqrt{\binom{2k}{k+m}\binom{2k}{k+m'}}}\\
&\times \binom{s}{r}\binom{n-k}{s}\binom{u}{v}\binom{j-k}{u}\binom{2k-l}{k-m-s+2r}\binom{2k-l}{k-m'-u+2v}\binom{2k}{l}\\
&\qquad\qquad\times \frac{\left(-\dfrac{q^2}{4}\right)^{\,n+j-k-r-v-l+\frac{1}{2}(m+m')}}{2^{s+u}\left(1+ \dfrac{q^2}{4}\right)^{\,n+j-l+\frac{1}{2}}},
\end{aligned}
\end{equation}
where $l_{\max}=\min(k+m+s-2r,k+m'+u-2v)$. In obtaining Eq.~\eqref{eq78} we use $q_1=q\sin\theta$ and   $q_2=q\cos\theta$.

Equation~\eqref{eq78} provides the explicit analytical form of the anisotropic exciton form factor. In contrast to the isotropic result~\eqref{eq59}, its derivation requires the full six-generator algebra \eqref{eq62} together with the additional $su(2)$ factorization of Subsec.~\ref{subsec:3.1}, reflecting the coupling between radial and angular degrees of freedom in the anisotropic problem. Although the corresponding interaction matrix elements were evaluated in our previous studies ~\cite{le2026prb, le2026cip}, the form factor itself was not isolated as an independent analytical result. Equation ~\eqref{eq78} therefore makes the underlying algebraic structure explicit and, being independent of the specific Rytova–Keldysh interaction, may also be useful for other two-dimensional quantum problems involving Fourier-space interactions.

\subsection{Numerical Validation of the Derived Analytical Excitonic Form Factors}
To independently validate the analytical form factors derived above, we performed direct numerical calculations of both $F_{nm;jm'}(t)$ in
Eq.~\eqref{eq59} and $G_{nm;jm'}(q,\theta)$ in Eq.~\eqref{eq78} for different combinations of the quantum numbers $n,m,j,$ and $m'$. The
operators $\hat r$, $\hat x$, and $\hat y$ were constructed directly from their original bosonic representations in Eq.~\eqref{eq46}, and the
matrix exponentials $e^{-t\hat r}$ and $e^{i(q_1\hat x+q_2\hat y)}$ were evaluated in truncated two-mode bosonic Fock spaces, without using the normal-ordered factorizations from which Eqs.~\eqref{eq59} and \eqref{eq78} were derived. For all tested states, including both diagonal and off-diagonal matrix elements, the numerical results systematically converge to the corresponding analytical values as the truncation order is increased.

\begin{figure}[ht]
\centering
\includegraphics[width=0.85\textwidth]{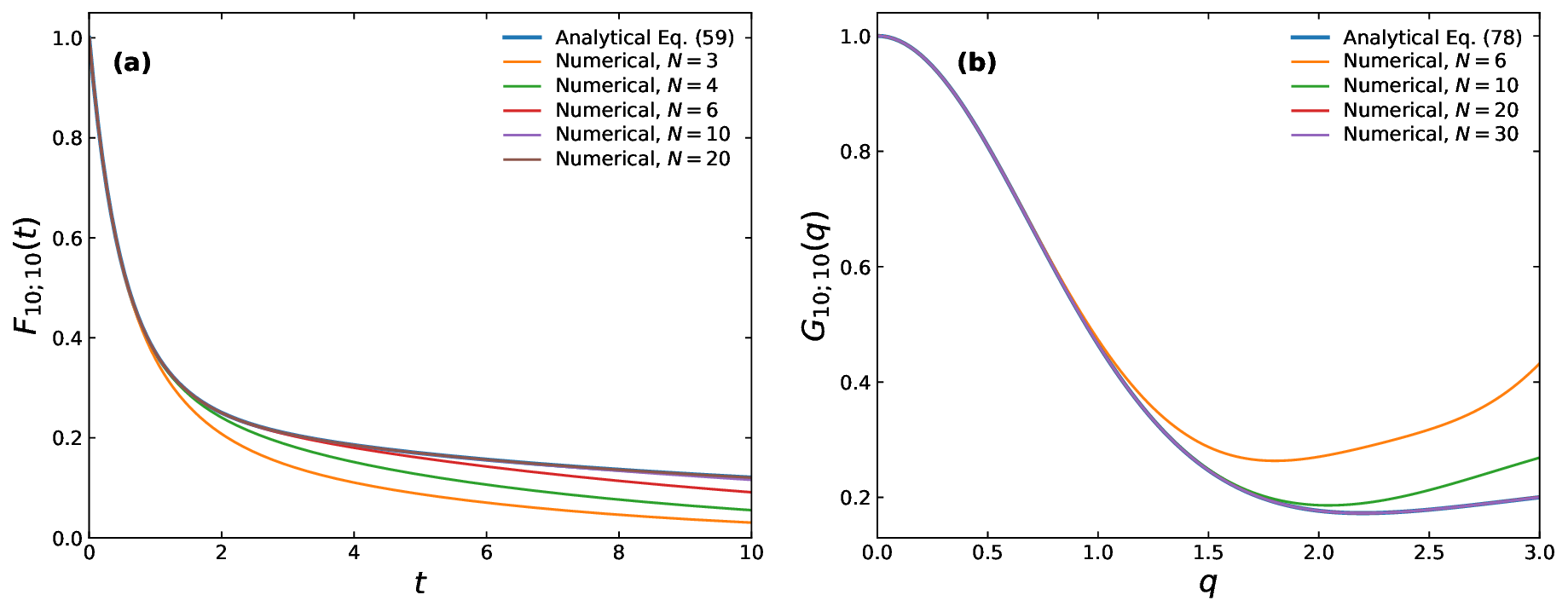}
\caption{Numerical validation of the analytical form factors.
(a) Comparison of $F_{10;10}(t)$ obtained from Eq.~\eqref{eq59}
with direct numerical calculations for truncation orders
$N=3,4,6,10,$ and $20$.
(b) Comparison of $G_{10;10}(q)$ obtained from Eq.~\eqref{eq78}
with direct numerical calculations for
$N=6,10,20,$ and $30$.}
\label{fig:benchmark}
\end{figure}

As a representative demonstration for Eq.~\eqref{eq59}, we consider $F_{10;10}(t)=\langle 1,0|e^{-t\hat r}|1,0\rangle$. Figure~\ref{fig:benchmark} compares the analytical result with direct numerical calculations for truncation orders $N=3,4,6,10,$ and $20$ over the interval $0\leq t\leq10$, while Table~\ref{tab:benchmark59} gives representative values and the maximum absolute deviation. The maximum deviation decreases from $9.03\times10^{-2}$ at $N=3$ to $1.96\times10^{-5}$ at $N=20$. The convergence becomes considerably slower for larger $t$. For example, at $t=30$, Eq.~\eqref{eq59} gives $F_{10;10}=0.0551758$, whereas truncations $N=10,20,30,$ and $50$ yield relative errors of approximately $47.4\%$, $7.5\%$, $0.97\%$, and $0.013\%$, respectively. This slow convergence at large $t$ illustrates the practical advantage of the closed analytical expression over a direct truncated-space evaluation.

\begin{table}[ht]
\centering
\caption{Numerical validation of the form factor $F_{10;10}(t)$ in
Eq.~\eqref{eq59}. The analytical values are compared with direct
calculations in truncated two-mode bosonic Fock spaces of order $N$.
The last column gives the maximum absolute deviation over
$0\leq t\leq 10$.}
\label{tab:benchmark59}
\small
\begin{tabular*}{\textwidth}{@{\extracolsep{\fill}}c|ccc|c}
\hline\hline
$N$ & $F^{(N)}(1)$ & $F^{(N)}(5)$ & $F^{(N)}(10)$
& $\max |\Delta|$ \\
\hline
3  & 0.35782779 & 0.08734168 & 0.03036085 & $9.03\times10^{-2}$ \\
4  & 0.36930654 & 0.12612493 & 0.05521226 & $6.52\times10^{-2}$ \\
6  & 0.37036553 & 0.15981177 & 0.09100690 & $2.94\times10^{-2}$ \\
10 & 0.37037037 & 0.16883818 & 0.11590985 & $4.46\times10^{-3}$ \\
20 & 0.37037037 & 0.16909620 & 0.12035075 & $1.96\times10^{-5}$ \\
\hline
Eq.~\eqref{eq59}
   & 0.37037037 & 0.16909621 & 0.12037037 & --- \\
\hline\hline
\end{tabular*}
\end{table}

For Eq.~\eqref{eq78}, we similarly illustrate the convergence using the diagonal matrix element $G_{10;10}(q)=\langle 1,0|e^{i\mathbf{q}\cdot\mathbf{r}}|1,0\rangle$. Figure~\ref{fig:benchmark} shows the analytical result together with direct numerical calculations for several truncation orders over $0\leq q\leq3$, and Table~\ref{tab:benchmark78} gives representative values. The maximum absolute deviation over this interval decreases from $2.32\times10^{-1}$ for $N=6$ to $1.34\times10^{-6}$ for $N=30$. We also tested off-diagonal matrix elements, which provide a more stringent check because both the magnitude and the complex phase must be reproduced. For example, for $(n,m;j,m')=(2,1;3,0)$,
$\theta=\pi/5$, and $q=2$, Eq.~\eqref{eq78} gives $G_{21;30}=0.0328420-0.0238611i$, while direct calculations with $N=20,30,$ and $40$ give absolute deviations of $7.20\times10^{-4}$, $8.48\times10^{-7}$, and $1.99\times10^{-12}$, respectively. These results confirm the analytical
form factor in Eq.~\eqref{eq78} and again show the progressively slower convergence of the direct truncated-space calculation as the momentum
increases.

\begin{table}[ht]
\centering
\caption{Numerical validation of the form factor $G_{10;10}(q)$ in
Eq.~\eqref{eq78}. The analytical values are compared with direct
calculations in truncated two-mode bosonic Fock spaces of order $N$.
The last column gives the maximum absolute deviation over
$0\leq q\leq 3$.}
\label{tab:benchmark78}

\small
\begin{tabular*}{\textwidth}{@{\extracolsep{\fill}}c|ccc|c}
\hline\hline
$N$ & $G^{(N)}(1)$ & $G^{(N)}(2)$ & $G^{(N)}(3)$
& $\max |\Delta|$ \\
\hline
6  & 0.47340940 & 0.27039535 & 0.43227629 & $2.32\times10^{-1}$ \\
10 & 0.46517161 & 0.18649713 & 0.26888927 & $6.87\times10^{-2}$ \\
20 & 0.46510214 & 0.17677038 & 0.20069606 & $4.79\times10^{-4}$ \\
30 & 0.46510214 & 0.17677670 & 0.20021601 & $1.34\times10^{-6}$ \\
\hline
Eq.~\eqref{eq78}
   & 0.46510214 & 0.17677670 & 0.20021723 & --- \\
\hline\hline
\end{tabular*}
\end{table}

\section{Conclusion}\label{sec:4}
In this work, we have developed a systematic algebraic approach to the normal ordering of exponential bosonic operators based on the Wei--Norman factorization method. By introducing an auxiliary parameter, the operator-ordering problem is transformed into a system of ordinary differential equations whose structure is determined by the commutation relations of the underlying closed Lie algebra. Starting from the Heisenberg--Weyl algebra for exponentials linear in bosonic operators, we extended the procedure to quadratic operators associated with the $su(1,1)$ algebra and subsequently to two-mode bosonic systems involving the $su(2)$ algebra and a six-generator closed algebra containing two coupled $su(1,1)$ subalgebras. These examples demonstrate that the same constructive procedure can be applied systematically as the algebraic structure of the exponential operator becomes increasingly involved.

As a physical application, we applied the approach to the analytical evaluation of excitonic form factors in two-dimensional semiconducting materials. The Levi--Civita transformation provides a natural bosonic formulation of the excitonic problem, while the Laplace and Fourier representations of the Rytova--Keldysh interaction reduce the corresponding interaction matrix elements to exponential bosonic form factors. For isotropic excitons, the Laplace representation leads to the form factor $\langle n,m|e^{-rt}|j,m'\rangle$, whose normal ordering is governed by a three-generator $su(1,1)$ algebra. For anisotropic excitons, the Fourier representation leads to $\langle n,m|e^{i{\bf q}\cdot{\bf r}}|j,m'\rangle$ and requires the full six-generator closed algebra, followed by an additional $su(2)$ factorization to complete the normal ordering in terms of the elementary bosonic operators. 
Explicit analytical expressions for both form factors have been obtained, making their underlying algebraic structures transparent and providing them as independent analytical results.

The present formulation separates the algebraic normal-ordering problem from the particular interaction for which the resulting form factors are employed. Consequently, the analytical expressions derived here are not restricted to the Rytova--Keldysh interaction: the isotropic form factor can be used in problems involving radial exponential operators or interactions admitting Laplace-type representations, while the anisotropic form factor may be applicable to other two-dimensional quantum problems formulated through Fourier-space interactions. More generally, the approach provides a direct and constructive route to normal-ordered representations of exponential bosonic operators whenever the relevant operators generate a closed Lie algebra.

\section*{Acknowledgements}
This research is funded by the Vietnam National Foundation for Science and Technology Development (NAFOSTED) under Grant No. 103.01-2023.138.

\section*{Author contributions}

Duy-Anh P. Nguyen: Investigation, Formal analysis. Ngoc-Tram D. Hoang: Investigation, Formal analysis. Dang-Khoa D. Le: Investigation, Formal analysis. Van-Hoang Le: Methodology, Investigation, Formal analysis, Funding acquisition, Supervision.

\section*{Conflict of interest}
The authors have no conflict of interest to declare.

\bibliographystyle{cip-v3-bst}
\bibliography{cip-v3-references}

\end{document}